\documentclass[twocolumn,prb,showpacs,floats,superscriptaddress]{revtex4}
\usepackage[dvips]{graphicx}
\usepackage{amsmath,bm}
\usepackage{psfrag}

\begin{document}

\title{Capture of particles of dust by convective flow}

\author{Dmitry V. Lyubimov}
\affiliation{Theoretical Physics Department, Perm State
University, Bukirev 15, 614990 Perm, Russia}
\author{Arthur V. Straube\footnote{E-mail: straube@stat.physik.uni-potsdam.de \\[0.5mm] {\mbox{} Paper published in Physics of Fluids {\bf{17}},
063302 (2005)}}} \affiliation{Department of Physics, University of
Potsdam, Am Neuen Palais 10, PF 601553, D-14415 Potsdam, Germany}
\affiliation{CFD Laboratory, Institute of Continuous Media
Mechanics, UB of Russian Academy of Sciences, Korolev 1, 614013
Perm, Russia}
\author{Tatyana P. Lyubimova}
\affiliation{Theoretical Physics Department, Perm State
University, Bukirev 15, 614990 Perm, Russia} \affiliation{CFD
Laboratory, Institute of Continuous Media Mechanics, UB of Russian
Academy of Sciences, Korolev 1, 614013 Perm, Russia}


\begin{abstract}
Interaction of particles of dust with vortex convective flows is
under theoretical consideration. It is assumed that the volume
fraction of solid phase is small, variations of density due to
nonuniform distribution of particles and those caused by
temperature nonisothermality of medium are comparable. Equations
for the description of thermal buoyancy convection of a dusty
medium are developed in the framework of the generalized
Boussinesq approximation taking into account finite velocity of
particle sedimentation. The capture of a cloud of dust particles
by a vortex convective flow is considered, general criterion for
the formation of such a cloud is obtained. The peculiarities of a
steady state in the form of a dust cloud and backward influence of
the solid phase on the carrier flow are studied in detail for a
vertical layer heated from the sidewalls. It is shown that in the
case, when this backward influence is essential, a hysteresis
behavior is possible. The stability analysis of the steady state
is performed. It turns out that there is a narrow range of
governing parameters, in which such a steady state is stable.
\end{abstract}



\maketitle

\section{Introduction}

Disperse systems, such as clouds of small particles suspended in
liquid or gas are widespread in natural environment as well as in
various fields of human activity. Due to admixture, such systems
demonstrate intriguing physical effects, which cannot occur in
homogeneous media. Understanding of the mechanisms governing the
behavior of particles in fluid flows is of crucial importance both
for fundamental studies and practical applications.

Advection of particles in fluid flows is currently investigated in
several disciplines. In chemistry (see, for example,
Refs.~\onlinecite{williams-85, peters-00}), this problem is
essential in lowering fuel consumption during combustion of liquid
fuels. In biology, recent developments have put forth a number of
fascinating problems concerning transport of microorganisms in
aqueous medium.\cite{koch-meinhardt-94, abraham-98,
delgiorgio-duarte-02} A typical example are algae, transported by
ocean flows. These algae contribute essentially to the absorption
of ${\rm CO}_2$ in oceans, which is closely related to the problem
of climate change. In rapidly developing microfluidics, advection
of particles has been the focus of close attention due to a need
of finding effective mechanisms of particle
mixing.\cite{ottino-89} In fluid mechanics light-scattering
particles of dust are widely used for visualization of flows.
Besides, it is also important to know how particles influence the
flow. Investigation of a spread of fine-dispersed impurities in
the atmosphere and oceans is tightly related to the problem of
environmental protection. These numerous applications have
motivated the unflagging interest in the fundamental problems of
particle advection in fluids.

The behavior of a small single particle in isothermal fluid flows
has been the subject of a great deal of study (see, for example,
the monograph \cite{soo-67} and the reviews \cite{marble-70,
maxey-90}). In particular, the equation of motion for a small
single particle in a nonuniform unsteady flow was discussed by
Maxey and Riley.\cite{maxey-riley-83} However, the fact that this
equation takes into account the integral Basset force essentially
complicates its solution. Therefore in most cases the problem has
been solved in terms of asymptotic theoretical models. Stommel
\cite{stommel-49} studied sedimentation of particles in a cellular
fluid flow in terms of weak inertia approximation. To our
knowledge this was the first paper that developed a simple
description of capture of small solid particles by a fluid flow.
It was shown, that when the ascending flow is rather intensive,
some particles are involved in a vortex motion and remain
suspended. Recently, capture of particles has been observed in a
cellular convective flow.\cite{simon-pomeau-91} Qualitatively
similar behavior is shown by granular media in the ascending air
flow under gravity: at a certain velocity of air flow the granular
matter is liquidized.\cite{anderson-etal-95} The problem of
sedimentation of small heavy particles has been generalized for
the case of a random flow (see Ref.~\onlinecite{pasquero-etal-03}
and references therein). However, a backward effect of the
particles on a fluid flow was not considered. A number of
theoretical and experimental studies are devoted to the problem of
particle accumulation in fluid flows -- the effect, which is
essential due to a difference in the inertial properties of the
phases when gravity is
insignificant.\cite{tio-etal-93,druzhinin-95-phf,
druzhinin-95-jfm,schwabe-frank-99} The situation, when both
gravity and inertia contribute to the particle dynamics, has been
studied in papers.\cite{maxey-corrsin-86, maxey-87,davila-hunt-01}
The authors of papers \cite{thomas-92, druzhinin-ostrovsky-94,
yanna-etal-97,mordant-pinton-00,
armenio-fiorotto-01,candelier-etal-04} have investigated the
influence of the Basset history force on the particle motion.

Another problem that has been intensively studied for many years
is the stability of laminar isothermal flows carrying small solid
particles.  First papers \cite{saffman-62,michael-64,liu-65} were
devoted to the stability of the particulate Poiseuille flow. This
problem has been generalized in the series of
papers.\cite{isakov-rudyak-95,rudyak-isakov-bord-97} The stability
of the Couette flow was considered in Ref.~\onlinecite{drew-75}. A
general result of these investigations is that stability of a
dusty flow qualitatively depends on the size of particles:
relatively small particles destabilize the flow, whereas the
particles of a larger size make the flow more stable. The
relaxation time of small particles is short, therefore such
particles move actually with the velocity of the fluid. The
presence of particles results in renormalization of the media
density. A larger efficient density is equivalent to a higher
fluid velocity, which leads to destabilization. On the contrary,
the relaxation time of large particles is long. Such particles are
too inert to respond to rapid velocity variations of the carrying
fluid and hence, large particles damp the fluid velocity
perturbations.

The collective behavior of particles in laminar nonisothermal
flows is much less studied. Moreover, the problem of interaction
of the particles with convective flows is not completely
understood yet. The exceptions are the works, concerning the
influence of settling particles on the stability of convective
flows.\cite{dementyev-96,dementyev-00,lyubimov-etal-98}

It is known, that particles of dust are widely used for
visualization of flows in the experiments on thermal buoyancy
convection. It is conventionally assumed that, if the mass
concentration of particles is small, their backward influence on
the convective flow is negligible. However, this is not always
true. The point is that buoyancy convective flow itself is caused
by small variations of density related to nonisothermality of a
medium. If density variations due to nonuniformity of the particle
distribution are of the same order of magnitude as those due to
nonisothermality, one should no longer ignore the influence of
particles on the flow. For small particle concentration this
effect is intuitively believed to be pronounced if the density of
particles and the density of the carrier fluid differ enough. A
typical example is solid particles suspended in a gaseous medium.
However, we argue below that the effect is essential in more
general case, when the densities of phases are different, but of
the same order (for example, in liquids laden with small solid
particles). Under gravity such particles tend to settle on the
bottom of a container. Two mechanisms hinder particles from
settling down: the Brownian motion of particles leading to their
diffusion in a carrier fluid and the entrapment of particles by
vortex fluid flows. The first mechanism operates also in a
quiescent medium and leads to the Boltzmann distribution of
particles over a height. However, for particle sizes commonly used
for flow visualization, an effective diffusion coefficient is so
small that in the equilibrium state the particles are practically
absent in most part of the volume. The second mechanism, which
operates only if the maximal velocity of ascending flows exceeds
the sedimentation velocity, is much more efficient. Investigation
of the capture effect and a backward influence of the particles on
a flow is important not only for visualization problems but also
for more fundamental aspects of environmental problems.

The mechanism of particle capture discussed above is related to
nonuniformity of the particle distribution over the volume. Even
if the initial distribution of particles is uniform, this
nonuniformity arises by itself due to the particle sedimentation
in the regions, where the flow is either descending or has small
vertical velocities. If the vertical dimension of the container is
large, the particle distribution in the most part of the container
is nearly uniform for a long time. However, even in this case the
particles can influence the flow, because their velocity differs
from the velocity of the fluid due to sedimentation. Thus, the
particles can transfer the perturbations of temperature and
vorticity and affect stability of a convective
flow.\cite{dementyev-96,dementyev-00}

In the present paper, we study the capture of dust particles by
convective flows. Particularly, a backward effect of particles on
the fluid flow is investigated in details. In
Sec.~\ref{sec:theor_model}, an appropriate theoretical model is
developed as a generalization of the Boussinesq approximation to a
convective flow in a dusty medium. In Sec.~\ref{sec:gen_consid},
the obtained equations are used to make a general interpretation
of the problem and to prove that the entrapped particles form a
cloud of dust. Section~\ref{sec:layer} is devoted to consideration
of a dusty medium in an infinite vertical layer. The existence of
a steady state solution with particle distribution in the form a
dust cloud is proved, and the stability of this state is
investigated.

\section{Theoretical model of thermal buoyancy convection in a dusty
medium \label{sec:theor_model}}
\subsection{Governing equations and basic assumptions}

Consider the behavior of small solid particles, suspended in a
nonisothermal fluid (liquid or gas) under gravity. On one hand,
all particles are assumed to be monodisperse spheres of a radius
$r_p$, which is large enough to neglect particle diffusion. On the
other hand, this size is supposed to be much smaller than the
characteristic length scale $L$ of the flow. On a scale much
greater than $r_p$ the particles are regarded as a continuous
medium with the volume fraction $\varphi=4/3\pi r_p^{3} n$ (the
volume fraction of the fluid phase $\mu=1-\varphi$), where $n$ is
the the number of particles per unit volume of the medium. Since
the actual volume concentration $n$ is proportional to $\varphi$
and differs from the latter by a constant factor, in the following
for the sake of simplicity $\varphi$ will be called concentration.

We suppose that the volume concentration of particles is small, so
we can neglect interparticle interactions and interactions between
the particles and walls of a container. It is also assumed that
the carrier phase is incompressible and the solid particles can
neither deform nor combine into agglomerates. Moreover, the
density of a solid phase is considered to be constant. The latter
assumption implies that we neglect thermal expansion of the solid
phase, which is perfectly valid in the dilute limit. After
averaging over space the equations for mass, momentum and energy
balance of both phases are written as follows: \cite{soo-67,
marble-70, nigmatulin-91}
\begin{subequations}
\label{ineqs-mass}
\begin{eqnarray}
\frac{\partial \left(\mu \rho \right)}{\partial t}+ {\rm
div}\left( \mu \rho {\bf u} \right) & = & 0, \\
\frac{\partial \varphi}{\partial t} +{\rm div}\left( \varphi {\bf
u}_p\right) & = & 0,
\end{eqnarray}
\end{subequations}
\vspace{-7mm}
\begin{subequations}
\label{ineqs-mom}
\begin{eqnarray}
\mu \rho\frac{D \bf u}{D t} & = & - \nabla p + \nabla \cdot
\left(\mu {\bf \varepsilon}\right) - \mu\rho {\rm
Ga}\,{\bf e}_z-\varphi{\bf F}, \\
\delta\, \frac{d{\bf u}_p}{d t} & = & -\delta\,{\rm Ga}{\bf
e}_z+{\bf F},
\end{eqnarray}
\end{subequations}
\vspace{-7mm}
\begin{subequations}
\label{ineqs-enrg}
\begin{eqnarray}
\mu \rho\frac{D T} {D t} & = & \frac{1}{\rm Pr}
{\rm div} \left(\mu \nabla T\right) + \frac{3}{\rm
Pr}\frac{L^2}{r_{p}^2}\varphi \left( T_p-T\right), \label{ineqs-enrg-a} \\
\delta\, {\rm B} \frac{d T_p} {d t} & = & - \frac{3}{\rm
Pr}\frac{L^2}{r_{p}^2} \left( T_p-T\right), \label{ineqs-enrg-b}
\end{eqnarray}
\end{subequations}
\noindent where ${\bf u}$ and ${\bf u}_p$, $T$ and $T_p$, are the
velocities and the temperatures of phases, $p$ and $\rho$ are the
pressure and the density of fluid, respectively (hereafter, the
subscript $``p"$ stands for the particle phase), ${\bf e}_z$ is
the unit vector of the axis $z$, directed against gravity; the
shear rate tensor $\varepsilon_{ij} = \nabla_i u_j+ \nabla_j u_i$.
The dimensionless variables have been introduced using the
following units: the reference density of fluid $\rho_0$ for the
densities of phases, the reference temperature difference $\theta$
for the temperatures, $L$ for the coordinates, $L^{2}/\nu$ for the
time, $\nu /L$ for the velocities, $\rho_0 \nu^{2} / L^{2}$ for
the pressure, where $\eta$ is the viscosity of fluid and $\nu=\eta
/\rho_0$.

Equations (\ref{ineqs-mass})-(\ref{ineqs-enrg}) involve the
following dimensionless parameters: the Galilei number ${\rm
Ga}=gL^3/\nu^2$, the Prandtl number ${\rm Pr}= \eta c/ \kappa$,
the ratio of densities of phases $\delta=\rho_p/\rho_0$, the ratio
of the specific heats of phases at constant pressure ${\rm
B}=c_p/c$ and the ratio of characteristic particle size to the
flow length scale $r_p/L$, where $\kappa$ is the thermal
conductivity of fluid. We make a note of the distinction between
the two Lagrangian derivatives $D/Dt=\partial /\partial t+{\bf u}
\cdot \nabla$ and $d/dt=\partial /\partial t+{\bf u}_p \cdot
\nabla$, which are used to denote the time derivative associated
with the motion of the fluid element and the element of a solid
phase, respectively.

In the interphase interaction term, ${\bf F}$ has the meaning of
the force exerted by an unsteady nonuniform fluid on the solid
particle. This force can be written as
\cite{maxey-riley-83,nigmatulin-91}
\begin{eqnarray} \label{ref11-3}
{\bf F} &=& \rho\left( \frac{D \bf u}{D t} + {\rm Ga}{\bf
e}_z\right)-\frac{9}{2}\frac{L^{2}}{r_p^2}{\bf W} \nonumber \\
&&-\frac{9}{2}\frac{L}{r_p}\sqrt{\frac{\rho}{\pi}}\int\limits_0^t
\frac{d {\bf W}(\tau)}{d \tau}
\frac{d\tau}{\sqrt{t-\tau}}-\frac{1}{2}\rho\frac{d{\bf W}}{d t},
\end{eqnarray}
\noindent where the relative velocity of phases ${\bf W}={\bf
u}_p-{\bf u}$ is introduced. The first two terms in
(\ref{ref11-3}), correspond to contributions to the force exerted
on the particle by undisturbed fluid flow due to gravity and
pressure gradient. In the approximation of undisturbed flow these
terms coincide with the Archemedian force. Generally, the particle
during its motion, disturbs the flow, the effect being taken into
account by the next three terms in (\ref{ref11-3}): the Stokes
viscous drag, the Basset history force, caused by unsteadiness of
a viscous boundary layer around a particle, and the added mass
force, allowing for the inertia of the surrounding fluid.

We do not consider the effects that can be initiated by rotation
of particles. Particularly, the Magnus force is not taken into
account in (\ref{ref11-3}). This assumption holds only in the case
when the particle response time $2{r_p^2}\delta/(9\nu)$ is much
less than the characteristic hydrodynamic time scale $L^{2}/\nu$.
Therefore the Stokes number, corresponding to a ratio of these
time scales, is assumed to be small:
$$
{\rm St} = \frac{2}{9} \frac{r_p^2}{L^2} \delta \ll 1.
$$
\noindent However, we assume the parameter $\delta$ to be finite
and do not restrict our theory to the limiting case of dusty media
($\delta \gg 1$), cf. Refs \onlinecite{druzhinin-95-phf,
druzhinin-95-jfm}. Further, we do not explicitly take into account
Einstein's correction term to the viscosity of fluid due to solid
admixture, which is inessential in the dilute limit.

The equation of state used here is typical for thermal buoyancy
convection. We consider weak convection, when the variations of
density due to nonisothermality of a fluid are small. Assuming
that the fluid density is a function of only temperature
$\rho=\rho(T)$, we expand it into a Taylor series near its
reference value $\rho_0$ at the temperature value $T_0$ and
restrict ourselves to a linear approximation
$$
\rho=1-\beta\theta T, \qquad \beta\theta \ll 1,
$$
\noindent where temperature $T$ is measured from the value $T_0$,
$\beta$ is the thermal expansion coefficient.

\subsection{Single-fluid approximation}

Let us consider a fluid laden with small solid particles and prove
that the initial two-phase theoretical model
(\ref{ineqs-mass})-(\ref{ref11-3}) can be simplified. We assume
that the mass concentration of particles $\delta\varphi$ is
comparable with the relative variations of the fluid density due
to nonisothermality. Bearing in mind that the particle
concentration $\varphi$, the nonisothermality $\beta\theta$, and
the relative size of particles $r_{p}/L$ are small we retain only
the leading terms in the momentum equations (\ref{ineqs-mom}). As
a result we conclude that in the basic state the pressure
distribution is hydrostatic, and obtain a relation between the
velocities of phases:
\begin{eqnarray}
\nabla p + {\rm Ga} {\bf e}_{z} & = & 0, \label{ref12-1} \\
{\bf u}_{p} = {\bf u} - {\rm S} {\bf e}_{z}, \qquad {\rm S} & = &
\frac{2}{9}\frac{r_{p}^{2}}{L^{2}}(\delta-1) {\rm Ga}
\label{ref12-2},
\end{eqnarray}
\noindent where $\rm S$ is the dimensionless sedimentation
velocity. Equation (\ref{ref12-2}) indicates that in the accepted
approximation the particle velocity equals the velocity of the
fluid plus the constant sedimentation velocity.

Taking into account (\ref{ref12-2}), we obtain in the next order
the momentum equation of the fluid
\begin{equation} \label{ref12-4}
\frac{\partial {\bf u}}{\partial t} + {\bf u} \cdot \nabla {\bf u}
= - \nabla p^{\prime} + {\nabla}^2 {\bf u} + \left[ \beta \theta
\, T - (\delta-1)\varphi \right ] {\rm Ga} {\bf e}_{z},
\end{equation}
\noindent where $p^{\prime}$ is the convective addition to the
hydrostatic pressure. Following the idea of the Boussinesq
approximation we can state that at large values of the Galilei
number and small density inhomogeneities  the density of the
medium will be constant everywhere except for the buoyancy force,
where these small density variations are multiplied by ${\rm Ga}$.
So, we introduce the thermal Grashof number ${\rm Gr}=\beta \theta
{\rm Ga}$ and its concentration analog ${\rm
Gc}=(\delta-1)\varphi_{0} {\rm Ga}$, where $\varphi_0$ is the
characteristic value of the particle concentration.
%

From the energy balance equation (\ref{ineqs-enrg-b}) we obtain in
the leading order: $T_{p} = T$. We restrict our consideration to
this approximation. This implies that for small particles the time
to equilibrate the phase temperature is much less than the viscous
hydrodynamic time scale. The energy equation for the fluid phase
is then obtained from Eqs. (\ref{ineqs-enrg}).
%

Taking into account the mass balance equations for fluid and
particles (\ref{ineqs-mass}) we arrive at a complete set of
equations describing thermal buoyancy convection in a
nonisothermal fluid laden with particles:
\begin{eqnarray}
\frac{\partial {\bf u}}{\partial t} + {\bf u} \cdot \nabla {\bf u}
& = & - \nabla p + {\nabla}^2 {\bf u} + \left({\rm Gr} \, T -
{\rm Gc} \varphi \right){\bf e}_{z}, \label{ref12-7a} \\
\frac{\partial T}{\partial t} + {\bf u} \cdot \nabla T & = &
\frac{1}{{\rm Pr}} {\nabla}^2 T, \label{ref12-7b} \\
\frac{\partial \varphi}{\partial t} + {\bf u}_{p} \cdot \nabla
\varphi & =& 0, \label{ref12-7c} \\
{\rm div}\,{\bf u} & =& 0, \quad {\bf u}_{p} = {\bf u} - {\rm
S}{\bf e}_{z}, \label{ref12-7d}
\end{eqnarray}

\noindent where the prime for $p$ is omitted, and the
concentration is normalized by $\varphi_{0}$.

The set of equations (\ref{ref12-7a})-(\ref{ref12-7d}) is
reminiscent of equations for thermosolutal convection. However,
there are two essential differences: first, we have neglected the
diffusion of particles, second, we have taken into account the
finite sedimentation velocity. The developed model is quite
general and can be applied to dusty media, aerosols, liquids laden
with small solid particles, and biological species in aqueous
media.

\section{Capture of dust particles. General consideration \label{sec:gen_consid}}

Consider the behavior of a dusty medium in a closed cavity of an
arbitrary shape heated from the sidewalls. It is assumed that at
the initial time the dust is uniformly distributed all over the
volume and at a later time no new particles enter the cavity but
the existing particles may settle to the bottom. We also assume
that the particles that have settled on the bottom are not carried
back into the flow, but stick to the bottom. The question arises
as to whether all particles should eventually settle on the bottom
or there may occur a situation when some particles will be
entrapped by the flow  and stay in a suspended state for
infinitely long time.

Obviously, in the absence of heating (${\rm Gr}=0$) and any
initial perturbations, all particles will eventually sink down on
the bottom for a time of order $1/{\rm S}$, which corresponds to
the following solution of the equations
(\ref{ref12-7a})-(\ref{ref12-7d}):
\begin{eqnarray*}
\label{ref2-1}
{\bf u} & = & 0, \\
      T & = & 0, \\
\varphi & = & \begin{cases}
0, & z > Z(t) \\
1, & z < Z(t) \end{cases}
\end{eqnarray*}

\noindent where $Z(t)={\rm const}-{\rm S}\,t$. So, initial
distribution of particles ensures sedimentation without distortion
of the interface between a pure fluid and suspension.

In order to simplify further argumentation, we will measure $z$
from the lowest point of the cavity. It can be understood through
insight that inducing of initial perturbations of concentration
and velocity does not change the situation: finally all
perturbations will be damped and all the particles will settle on
the bottom. To demonstrate this let us consider the evolution of
the total energy of the system $E$:
\begin{equation} \label{ref2-2}
E= \int_\Omega \left( \frac{u^2}{2} + {\rm Gc} \,\varphi z \right)
d\Omega.
\end{equation}

Using (\ref{ref12-7a})-(\ref{ref12-7d}) we obtain for the rate of
the energy change
\begin{equation} \label{ref2-3}
\frac {dE}{dt}= -\int_\Omega {\left( \nabla\times\, {\bf u}
\right)}^2 d\Omega + \oint_F \varphi z \, {\bf e}_z\cdot{\bf
n}\,dF,
\end{equation}

\noindent where $F$ is the surface bounding the volume $\Omega$,
${\bf n}$ is the external vector normal to $F$. To obtain
Eq.~(\ref{ref2-3}) the no slip conditions for fluid velocity are
assumed: ${\bf u} \vert_F =0$. Since at all time moments $t>0$ the
value $\varphi$ differs from zero only on such areas of the
boundary surface where ${\bf e}_z \cdot {\bf n}<0$ (i.e. at the
bottom of a cavity), the right-hand side of (\ref{ref2-3}) is
always nonpositive. Moreover it can be equal to zero only in the
case of a quiescent fluid in the absence of the particles at the
walls. This proves that the energy $E$ can only decrease. On the
other hand, as can be seen from (\ref{ref2-2}), $E \ge 0$, i.e. it
is bounded from below. Hence, the final state is that, at which
the right-hand side in (\ref{ref2-3}) turns to zero, i.e. when the
fluid becomes completely free of the particles and all fluid flows
cease.

The situation radically changes if there is a permanent external
source that makes fluid move. In the present work, the source of
this kind is  side heating of the wall, generating a vortex
convective flow in a cavity. Naturally, in this case, the areas
near the walls will finally be free of particles. Indeed, as it
follows from Eq.~(\ref{ref12-7c})
\begin{equation} \label{ref2-4}
\frac {d}{dt} \int_\Omega \varphi \, d\Omega = -\int_\Omega
 {\bf u}_{p} \cdot \nabla \varphi \, d\Omega = {\rm S} \oint_F \varphi
\, {\bf e}_z \cdot {\bf n} \, dF \le 0.
\end{equation}

\begin{figure}[!t]
\includegraphics[width=0.45\textwidth]{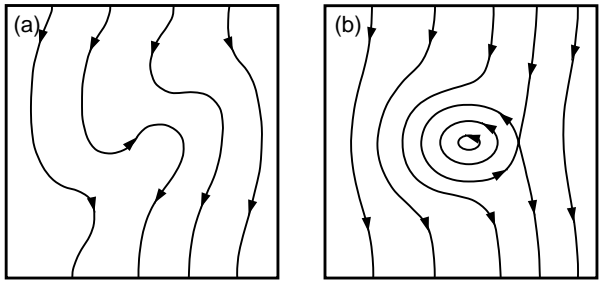}
\caption{Streamlines of the particle flow in the absence (a) and
in the case (b) of particle capture.} \label{fig:capt_sketch}
\end{figure}

\noindent The total amount of the dust in the cavity decreases
until $\varphi$ turns to zero everywhere near the boundaries. It
does not mean, however, that the bulk of the fluid will be
completely free of particles. If in some part of the cavity the
velocity of the ascending fluid $u_a$ exceeds the sedimentation
velocity ${\rm S}$, we can expect the capture of the particles.
The condition
\begin{equation} \label{ref2-5}
u_a > {\rm S}
\end{equation}
\noindent is the necessary condition for capture but it is not a
sufficient one. Indeed, we can easily imagine the flow pattern
with the streamlines of particles shown in
Fig.~\ref{fig:capt_sketch}(a): despite the existence of regions
with ascending motion of particles, all streamlines end at the
lower boundary.
%
%
%
%

On the other hand, we can specify a sufficient condition for
capture, which is probably not necessary. Let us assume that there
is the line, at which the horizontal component of the the fluid
velocity vanishes. If on this line there is a point where $u_a =
{\rm S}$ then such a point is a singular point for the vector
field ${\bf u}_{p}$. Furthermore, owing to (\ref{ref12-7d}) we
have
$$
{\rm div} \, {\bf u}_{p} =0,
$$
\noindent and therefore the vector field ${\bf u}_{p}$ is free of
sources or sinks. In this case only two types of structurally
stable fixed points are possible: center or saddle. In the case of
the center point, there are closed orbits, i.e. some particles are
entrapped by a vortex and stay inside for infinitely long time.
The existence of the saddle point together with the boundary
conditions for ${\bf u}_{p}$ necessary implies the existence of
the closed separatrix loop and the center point, as it is
schematically presented in Fig.~\ref{fig:capt_sketch}(b). The
separatrix loop forms the boundary of the captured cloud of
particles.

\begin{figure}[!b]
\includegraphics[width=0.45\textwidth]{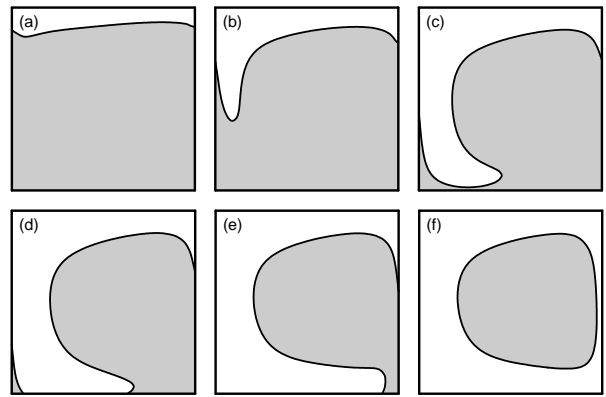}
\caption{Evolution of particle concentration $\varphi$ to the
steady state at ${\rm Gr}=10~000$, ${\rm S}=5$, ${\rm Gc}=0$. The
times for the states of $\varphi$ are: (a) $t=0.02$, (b) $t=0.04$,
(c) $t=0.07$, (d) $t=0.09$, (e) $t=0.14$, (f) $t=0.40$. The
regions of pure liquid and particle suspension are shown in white
and grey respectively.} \label{fig:formation}
\end{figure}

To demonstrate how formation of a cloud of dust occurs, we have
numerically integrated Eqs.~(\ref{ref12-7a})-(\ref{ref12-7d}) for
a partial case of the square cavity, heated from the right side.
The quiescent state with uniform distribution of particles was
chosen as the initial one. The values of governing parameters
correspond to the single-vortex flow, satisfying the condition
(\ref{ref2-5}). This complementary example is in agreement with
the results of general consideration. During evolution (see
Fig.~\ref{fig:formation}), a part of particles is gradually
leaving the flow, whereas other particles stay suspended. As a
result of this transient process, the system evolves to the steady
state with a distribution of particles in the form of a cloud.
%
%
%
%

The theory above describes the case of no particle inertia. In
general, the trajectory of a captured particle is no longer a
closed orbit. Due to inertia the center fixed point [see
Fig.~\ref{fig:capt_sketch}(b)] transforms to a focus, which is
unstable for heavy particles.\cite{maxey-90} In the vicinity of
the focus, captured particles of dust are repelled by this fixed
point and move along spirals. Nonetheless, for small particles
this effect is weak. The addition to the velocity of particle due
to purely inertial drift ${\bf u}_p^{in}$ is given
\cite{druzhinin-95-phf,druzhinin-95-jfm} by the term ${\rm St}\,{D
\bf u}/{D t}$. Since the intensity of thermal convection is
governed by the thermal Grashof number, we can estimate ${\bf
u}_p^{in} \sim {\rm St}\,{\rm Gr}$. For fine particles of dust
with $r_{p} \sim 10^{-4}~\rm cm$ suspended in air on laboratory
length scales $L \sim 1~\rm cm$ at moderate intensity of thermal
convection we obtain the inertial time scale $\tau_{in} \sim
10^3~\rm s$ (${\rm St}\sim 10^{-5}$, ${\rm Gr}\sim 10^{3}$).

Recall, that we have also neglected the diffusion of the
particles, which means that our consideration concerns the time
scales less than the diffusion time scale $t_d$. However, this
restriction is practically of no significance. Indeed, according
to the Einstein formula, the diffusion coefficient of the
spherical particles $D=kT/(6\pi \eta r_{p})$ (here $T$ is the
absolute temperature, $k$ is the Boltzmann constant). As before,
for the fine particles of dust in air at room temperature we
obtain $D \sim 10^{-7}~\rm{cm^2/s}$, which for the laboratory
length scales corresponds to the diffusion time scale $t_d \sim
10^{7}~\rm s$ ($r_{p} \sim 10^{-4}~\rm cm$, $L \sim 1~\rm cm$).

Thus, despite the fact that existence of a state with a cloud of
dust is in general not everlasting due to particle inertia and
particle diffusion, it exists long enough to be important.

\section{Cloud of dust in a vertical layer \label{sec:layer}}
\subsection{Basic steady state}

A rather complete understanding of the mechanisms responsible for
formation of a cloud of dust and its backward influence on the
flow hydrodynamics can be obtained from a model problem. Let a
dusty medium fill the infinite vertical layer $-1 < x < 1$, the
boundaries of which are kept at constant, but different
temperatures: $T=-1$ at $x=-1$ and $T=1$ at $x=1$. We assume that
initial distribution of particles is uniform and look for a steady
state solution, in which the velocity has only vertical component
$u_0$ and all fields except for the pressure are independent of
the vertical coordinate $z$. Then, the set of equations
(\ref{ref12-7a})-(\ref{ref12-7d}) takes the form:
\begin{eqnarray}
u_0^{\prime \prime} + {\rm Gr}\, T_0 - {\rm Gc} \varphi_0 & = & c,
\label{ref31-1} \\
T_0^{\prime \prime} & = & 0,  \label{ref31-2}
\end{eqnarray}
\noindent where the prime stands for differentiation with respect
to $x$, the subscript $``0"$ is used to indicate the steady state
solution, $c=\partial p/\partial z =\rm{const}$. Equations
(\ref{ref31-1}), (\ref{ref31-2}) should be supplemented with the
boundary conditions for the fluid velocity and the temperature:
\begin{equation}
x=\pm 1: \quad u_0 = 0, \quad T_0=\pm 1. \label{ref31-3} \\
\end{equation}
In the case of the infinite layer we should also specify integral
conditions. We prescribe no-flux for fluid and particles, which
mean that the flow is closed at the infinity and no new particles
enter the system:
\begin{equation}
\int\limits_{-1}^1 u_0 \,dx =0,  \quad  \int\limits_{-1}^1
\varphi_0 \, u_{p\,0} \,dx=0, \label{ref31-5}
\end{equation}
\noindent where $u_{p\,0}=u_0-{\rm S}$.

According to (\ref{ref31-2}), (\ref{ref31-3}), the temperature
distribution does not depend on the fluid and particle motions and
can be obtained directly
\begin{equation} \label{ref31-7}
T_0=x.
\end{equation}

Let us analyze velocity profiles of the fluid and solid phase. We
start the discussion with the simplest case $\rm Gc=0$, when the
particles do not influence the fluid motion. In this case, the
velocity profile of the fluid is exactly the same as in the
well-known problem on a convective flow of pure fluid in a
vertical layer heated from the
sidewalls:\cite{gershuni-zhuhovitsky-76}
\begin{equation} \label{ref31-8}
u_0= \frac{{\rm Gr}}{6}\, x\left(1-x^2\right).
\end{equation}

The particles occupy the region of the layer between $x_1$ and
$x_2$. Within this interval $\varphi_0=1$, whereas in the
near-wall regions $\varphi_0=0$.

The point $x_2$ is the point at which the velocity of particle
sedimentation coincides with the velocity of the ascending fluid
flow
\begin{equation} \label{ref31-9}
x=x_2: \quad u_0={\rm S}.
\end{equation}
\noindent According to (\ref{ref31-8}), it is determined by the
largest root of the cubic equation
\begin{equation} \label{ref31-10}
{\rm Gr}\,x_2\left(1-x_2^2\right) = 6 {\rm S}.
\end{equation}

The point $x_1$ is determined by the zero-flux condition
(\ref{ref31-5}) for particles. It is convenient to rewrite the
equation for $x_1$ in terms of semi-width of the cloud
$d=(x_2-x_1)/2$ (hereafter, just width). Thus, we obtain for $d$
\begin{equation} \label{ref31-11}
d=x_2-\sqrt{\frac{1-x_2^2}{2}}.
\end{equation}

\begin{figure}[!t]
\includegraphics[width=0.35\textwidth]{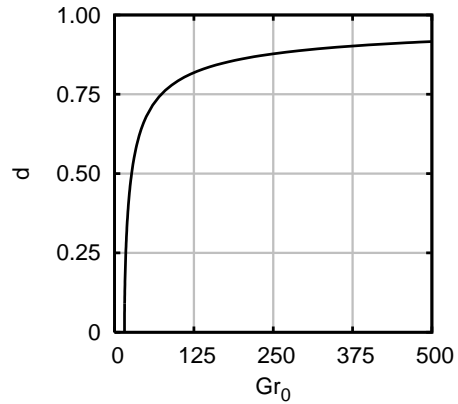}
\caption{The width of a dust cloud $d$ versus ${\rm Gr}_0$ at
${\rm Gc}=0$.} \label{fig:d_Gr}
\end{figure}

As evident from Eq.~(\ref{ref31-10}) and the second condition in
(\ref{ref31-5}), the parameter ${\rm S}$ is not independent; the
width of the cloud is determined by the ratio ${\rm Gr}_0={\rm
Gr/S}$, moreover, the cloud exists only if ${\rm Gr}_0 >
9\sqrt{3}$. With the growth of ${\rm Gr}_0$ the width of the cloud
monotonically increases and tends to $1$ as ${\rm Gr}_0\to\infty$
(see Fig.~\ref{fig:d_Gr}).
%
%
%

At nonzero values of ${\rm Gc}$ the particles exercise an
influence on the fluid flow, and the dependence of the
characteristics of the cloud on the parameters becomes more
complex. In this case the basic state is governed by ${\rm Gr}_0$
and a renormalized concentration parameter ${\rm Gc}_{0}={\rm
Gc/S}$. In the presence of a particle cloud the flow region splits
into three zones: 1) $-1<x<x_1$, 2) $x_1<x<x_2$, 3) $x_2<x<1$. In
each of these zones the velocity profile is described by
third-degree polynomials, which can be written down in the
following form, respectively:
\begin{eqnarray}\label{ref31-12}
u_0^{(1)} & = & u_0+c\left(1-x^2\right)+c_1\left(1+x\right), \nonumber \\
u_0^{(2)} & = & u_0+\frac{1}{2}\,{\rm
Gc}_0\,x^2+c\left(1-x^2\right)+c_{21}\,x+c_{22}, \\
u_0^{(3)} & = & u_0+c\left(1-x^2\right)+c_3\left(1-x\right),
\nonumber
\end{eqnarray}
\noindent where $u_0$ is given by (\ref{ref31-8}) normalized by
$\rm S$; the unknown coefficients of the polynomials $c_1$,
$c_{21}$, $c_{22}$, $c_3$, the parameter $c$ and the coordinates
of the points $x_1$ and $x_2$ are defined by the boundary
conditions (\ref{ref31-3}), (\ref{ref31-5}), the condition
(\ref{ref31-9}) and the continuity conditions for the velocity and
tangential stress at the cloud borders. The system of the obtained
algebraic equations is partly simplified: the constants $c$,
$c_1$, $c_{21}$, $c_{22}$, $c_3$ can be excluded by expressing
them  in terms of the width of the dust cloud $d$ and the
coordinate of its center $x_c=(x_2+x_1)/2$:
\begin{eqnarray*}
 c & = & d \, {\rm Gc}_0 \left [ 3\left(1-x_c^2\right)-{d}^2\right
 ], \\
c_1 & = & -d \, {\rm Gc}_0\left(1-x_c\right), \quad c_3=-d
\, {\rm Gc}_0 \left(1+x_c\right), \\
 c_{21} & =& -{\rm Gc}_0 \, x_c \,(1-d),
\quad c_{22}=\frac{1}{2}\, {\rm Gc}_0 \left(x_c^2+{d}^2-2d\right).
\end{eqnarray*}
The remaining unknowns $d$ and $x_c$ are defined by the couple of
nonlinear algebraic equations, which are solved numerically
\begin{eqnarray}
12 & = & 3 d \, {\rm Gc}_0\left(x_c+d-1\right) f_1(x_c,d)
\nonumber \\
&&\mbox{}-2 \, {\rm Gr}_0\left(x_c+d\right) \left [
\left(x_c+d\right)^2+1\right], \label{ref31-13a} \\
12 & = & 3 d \, {\rm Gc}_0 f_2(x_c, d)- 2 \, {\rm Gr}_0 x_c \left(
x_c^2+{d}^2-1\right), \label{ref31-13b}
\end{eqnarray}
\begin{eqnarray}
f_1(x_c,d) & = &
3x_c^3+3\left(d+1\right)x_c^2+\left(d^2+1\right)x_c \nonumber \\
&&\mbox{}+d^3+ d^2-3 d+1, \nonumber \\
f_2(x_c,d) & = & \left( 3x_c^2+d^2+2d-3 \right )\left(
3x_c^2+d^2-2d+1\right ). \nonumber
\end{eqnarray}
As in the case discussed above, we assume that $\varphi_0 =0$ in
the zones $1$ and $3$ and $\varphi_0 =1$ in the zone $2$.

The cloud is absent if the velocity of fluid is lower than the
velocity of sedimentation all over the volume. As in the case of
${\rm Gc}_0=0$, this means that the cloud can exist only at ${\rm
Gr}_0 > 9\sqrt{3}$.
%
%
\begin{figure}[!t]
\includegraphics[width=0.35\textwidth]{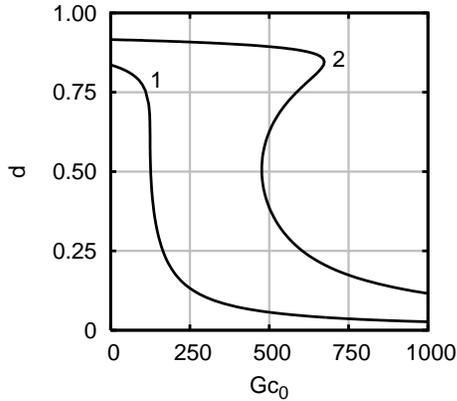}
\caption{The width of the cloud $d$ versus ${\rm Gc}_0$ at ${\rm
Gr}_0=150$ (line 1), ${\rm Gr}_0=500$ (line 2). }
\label{fig:hyster_d_Gc}
\end{figure}
%
%

Let us discuss the dependence of the width of the particle cloud
on the concentration parameter ${\rm Gc}_0$. At a relatively low
${\rm Gr}_0$ the width of the cloud $d$ monotonically decreases
with the growth of ${\rm Gc}_0$ (see Fig.~\ref{fig:hyster_d_Gc},
line 1). This is related to the fact that the major part of the
cloud is located in the ascending flow, hence, the larger ${\rm
Gc}_0$, the higher effective density of the medium in this place.
The higher density leads to a decrease of the buoyancy force and
therefore to lowering of the flow velocity. The latter, in its
turn, results in a decrease of the width of the dust cloud, in
which the particles can remain in the suspended state.

At higher values of ${\rm Gr}_0$ the dependence of $d$ on ${\rm
Gc}_0$ becomes more complex. Generally, in this case the growth of
${\rm Gc}$ also results in the decrease of the cloud width, but
now in some range of ${\rm Gc}_0$ the dependence $d({\rm Gc}_0)$
is no longer unique (see Fig.~\ref{fig:hyster_d_Gc}, line 2). At
low ${\rm Gc}_0$ the growth of ${\rm Gc}_0$ only slightly
influences the width of the cloud, but starting from some
threshold value of ${\rm Gc}_0$, the drag force of the wide cloud
becomes so strong that the flow is unable to keep it further. As a
result, the width of a cloud decreases by a jump. If now, starting
from a large value of ${\rm Gc}_0$, we decrease it, first, the
cloud width varies insignificantly, but then at some critical
value of concentration parameter the width of a cloud makes a jump
to nearly the highest possible value for a given ${\rm Gr}_0$.
Thus, the decrease of the cloud width with the change of particle
concentration demonstrates a well pronounced hysteresis.
%
%
\begin{figure}[!t]
\includegraphics[width=0.35\textwidth]{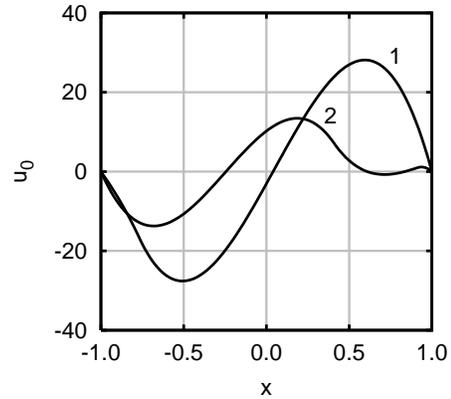}
\caption{Velocity profiles for upper (line 1) and lower (line 2)
branches of solutions at ${\rm Gr}_0=500$, ${\rm Gc}_0=600$. }
\label{fig:vprofiles}
\end{figure}
%

In Fig.~\ref{fig:vprofiles} the profiles of the fluid velocity at
${\rm Gr}_0=500$, ${\rm Gc}_0=600$ are plotted for upper (line 1)
and lower (line 2) branches of possible steady states. It is seen,
that for the upper branch the velocity profile only slightly
differs from the usual cubic profile (\ref{ref31-8}), but for the
lower branch the flow in the domain occupied by the cloud is
strongly suppressed, the total intensity of the flow is also low.

\begin{figure}[!t]
\includegraphics[width=0.35\textwidth]{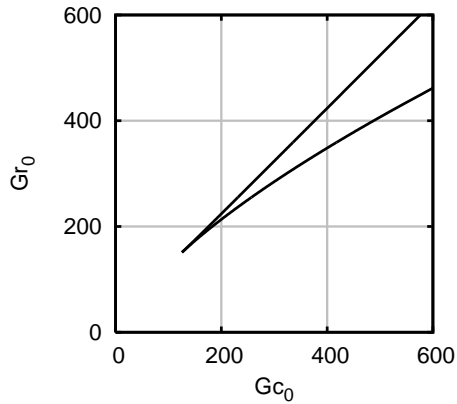}
\caption{Hysteresis range on the parameter plane (${\rm Gc}_0$,
${\rm Gr}_0$).} \label{fig:wedge_Gr_Gc}
\end{figure}

In Fig.~\ref{fig:wedge_Gr_Gc} we plot a diagram defining a range
of hysteresis existence on the plane ($\rm Gc_0$, $\rm Gr_0$). The
solid lines divide the plane into two parts: larger and smaller
ones. Any point of the larger part corresponds to only one
solution (one root of a cloud width $d$ at fixed $\rm Gc_0$, $\rm
Gr_0$), whereas in the smaller one, the hysteresis zone, there are
three solutions (see Fig.~\ref{fig:hyster_d_Gc}). The solid lines,
at which there exist two solutions, correspond to folds; the point
where these lines intersect is a cusp with one solution. Inside
the hysteresis zone, one of three solutions, namely, corresponding
to the middle branch of a hysteresis curve $d({\rm Gc}_0)$ in
Fig.~\ref{fig:hyster_d_Gc}, is structurally unstable.
%
%
%
%

\subsection{3.2. Linear stability analysis}
\subsubsection{Formulation of stability problem}

Let us investigate the stability of the basic steady state. We
restrict our consideration to a fixed value of the Prandtl number
${\rm Pr}=1$, which corresponds to a typical case of particles,
suspended in gaseous medium.

In order to formulate the stability problem, the governing
equations (\ref{ref12-7a})-(\ref{ref12-7d}) should be supplemented
by necessary conditions at the rigid boundaries of the layer and
continuity conditions at the borders of the dust cloud. At the
rigid walls of the layer we impose the no slip condition for fluid
velocity, and the condition of constant temperature:
\begin{equation}\label{ref321-1}
x=\pm 1: \quad {\bf u}=0, \quad T=\pm 1.
\end{equation}

The borders of the dust cloud, which are the surfaces of
discontinuity for particle concentration, must satisfy the
conditions of stress balance, the continuity of velocity,
temperature, and energy flux. These conditions are the direct
consequence of conservation laws for momentum, energy, and mass.
Assuming that the borders of a dust cloud are deformable, the
conditions at the interfaces are
\begin{eqnarray}
x=x_{1,2}+\zeta_{1,2}: && -[p]+\left[\varepsilon_{nn}\right]=0,
\; [\varepsilon_{n\tau}]=0, \nonumber \\
&&[{\bf u}]=0,  \; [T]=0, \; \left[\frac{\partial T}{\partial
n}\right]=0, \label{ref321-2}
\end{eqnarray}
\noindent where $\zeta_1=\zeta_1(y,z,t)$, $\zeta_2=\zeta_2(y,z,t)$
are the deviations of concentration surfaces from the flat
undisturbed shapes, and square brackets are used to denote the
jump of a function $[f]=f_2-f_{1,3}$. Here the subscripts $1$,
$2$, $3$ correspond to three zones of the flow: $1)$
$-1<x<x_1+\zeta_1$, $2)$ $x_1+\zeta_1<x<x_2+\zeta_2$, $3)$
$x_2+\zeta_2<x<1$. The unit normal $\bf n$ and tangential ${\bm
\tau}_1$, ${\bm \tau}_2$ vectors to the surface are given by the
relationships
\begin{eqnarray}
&&{\bf n}=\frac{{\bf
e}_x-\nabla\zeta_{1,2}}{\sqrt{1+\left(\nabla\zeta_{1,2}\right)^2}},\;\;
{ \tau}_1=\frac{{\bf e}_z+{\bf e}_x\,{\bf e}_z \cdot
\nabla\zeta_{1,2}}{\sqrt{1+\left({\bf e}_z \cdot
\nabla\zeta_{1,2}\right)^2}},\nonumber \\
&&{ \tau}_2={\bf n}\times { \tau}_1. \nonumber
\end{eqnarray}

\noindent Here we introduce the orts ${\bf e}_x=(1,0,0)$, ${\bf
e}_y=(0,1,0)$ of the axes $x$ and $y$, respectively. Both borders
of the dust cloud are impervious to the particles, the velocity of
any point of such border coincides with the Lagrangian velocity of
a particle at this point. Hence, for surface deviations the
following kinematic conditions are specified:
\begin{equation}\label{ref321-3}
\frac{\partial \zeta_{1,2}}{\partial t}+{\bf u}_{p}\cdot\nabla
\zeta_{1,2}={\bf u}_{p} \cdot {\bf e}_x.
\end{equation}

Let the basic steady state be disturbed by introducing small
perturbations of the velocity ${\bf v}$, pressure $q$, temperature
$\vartheta$ and concentration $\phi$. Then we substitute the
disturbed fields ${\bf u}_0+{\bf v}$, $p_0+q$, $T_0+\vartheta$,
$\varphi_0+\phi$ into Eqs.~(\ref{ref12-7a})-(\ref{ref12-7d}).
Neglecting squared and higher order terms with respect to
perturbations, we obtain the equations for perturbations:
\begin{eqnarray}
\frac{\partial {\bf v}}{\partial t}+{\bf u}_0\cdot\nabla{\bf
v}+{\bf v}\cdot\nabla{\bf u}_0 & = & -\nabla q+{\nabla}^2 {\bf v}
\mbox{}+{\rm Gr}\,\vartheta \,{\bf e}_z \nonumber \\
&&\mbox{}-{\rm Gc}\,\phi\,{\bf e}_z, \label{ref321-4a} \\
\frac{\partial \vartheta}{\partial t}+{\bf u}_0\cdot\nabla
\vartheta+{\bf v}\cdot\nabla T_0 & = & \frac{1}{\rm Pr}{\nabla}^2
\vartheta, \quad {\rm
div}\,{\bf v}=0,\label{ref321-4b} \\
\frac{\partial \phi}{\partial t}+\left( {\bf u}_0 -{\rm S}\,{\bf
e}_z \right)\cdot\nabla \phi & = & 0, \label{ref321-4c}
\end{eqnarray}
\noindent where ${\bf u}_0=\left(0,0,u_0(x)\right)$.

The boundary conditions (\ref{ref321-1}) take the form for
perturbations:
\begin{equation}\label{ref321-5}
x=\pm 1: \quad {\bf v}=0, \quad \vartheta=0.
\end{equation}

Let us also assume the smallness of $\zeta_1$, $\zeta_2$. Then,
the equations (\ref{ref321-4a})-(\ref{ref321-4c}), boundary
conditions (\ref{ref321-5}) and conditions at deformable
interfaces (\ref{ref321-2}) reduced to those for undisturbed
interfaces admit a transformation, which is analogous to the
Squire transformation \cite{squire-33} and discussed below. Such a
transformation allows us to reduce the full three-dimensional
problem to a problem in two dimensions. We denote the $x$ and $z$
components of the velocity $\bf v$ by $u$ and $w$, respectively
and analyze the stability of the basic steady state with respect
to the two-dimensional perturbations in the form of transversal
rolls. Thus, the solution can be written as normal modes
\begin{equation}\label{ref321-6}
\left( \begin{array}{cccccc} u(x,z,t) \\ w(x,z,t) \\ q(x,z,t) \\ \vartheta(x,z,t) \\
\phi(x,z,t) \\ \zeta_{1,2}(z,t)
\end{array} \right) = \left( \begin{array}{cccccc} \hat{u}(x) \\ \hat{w}(x) \\ \hat{q}(x) \\ \hat{\vartheta}(x) \\
\hat{\phi}(x) \\ \hat{\zeta}_{1,2}
\end{array}\right) e^{\lambda t - ikz},
\end{equation}

\noindent where $\lambda=\lambda_r+i\lambda_i$ is the complex
growth rate, $k$ is the wave number, characterizing periodicity of
perturbations along the $z$-axis.

Let us substitute the ansatz (\ref{ref321-6}) into
Eqs.~(\ref{ref321-4a})-(\ref{ref321-4c}). It follows from
(\ref{ref321-4c}) that
$$
\left\{\lambda-ik(u_0-{\rm S})\right\}\hat{\phi}=0,
$$

\noindent which must be satisfied at every point of a layer
occupied by particles. Since $u_0$ is a function of $x$ we assume
that $\left\{\lambda-ik(u_0-{\rm S})\right\}\ne 0$,
${\hat\phi}=0$. As a result, in each of the three zones the
problem is defined by the same set of amplitude equations:
\begin{eqnarray}
\lambda \hat{u} - ik u_0 \hat{u} & = & -\hat{q}'+\hat{u}''-k^2\hat{u}, \label{ref321-7a} \\
\lambda \hat{w} - ik u_0 \hat{u}+u_0'\hat{u} & = &
ik\hat{q}+\hat{w}''-k^2\hat{w}+{\rm Gr}\hat{\vartheta},
\label{ref321-7b} \\
\hat{u}'-ik\hat{w} & = & 0, \label{ref321-7c} \\
\lambda\hat{\vartheta} - ik u_0\hat{\vartheta}+\hat{u} & =&
\frac{1}{\rm Pr}\left(\hat{\vartheta}''-k^2\hat{\vartheta}\right).
\label{ref321-7d}
\end{eqnarray}

After linearization, the boundary conditions (\ref{ref321-5}) and
the conditions (\ref{ref321-2}) at the pure liquid-suspension
interface, i.e. at the borders of a dust cloud, written for the
amplitudes of perturbations, take the form:
\begin{eqnarray}
x =  \pm 1: && \hat{u}=0, \;\; \hat{w}=0, \;\;
\hat{\vartheta}=0, \label{ref321-7e} \\
x =  x_{1,2}: && [\hat{q}]=0, \;\; [\hat{u}]=0, \;\;
[\hat{w}]=0, \nonumber \\
&&[\hat{w}']=-{\rm Gc}\,\hat{\zeta}_{1,2}, \;\;
[\hat{\vartheta}]=0, \;\; [\hat{\vartheta}']=0, \label{ref321-7f}
\end{eqnarray}
\noindent where $\hat{\zeta}_1$ and $\hat{\zeta}_2$ are obtained
from Eqs.~(\ref{ref321-3}) together with (\ref{ref321-7f}) and are
determined by the relations
\begin{equation}\label{ref321-7g}
x=x_1: \;\; \left\{\lambda -ik(u_0-{\rm
S})\right\}\hat{\zeta}_1=\hat{u}; \quad x=x_2: \;\;
\lambda\hat{\zeta}_2=\hat{u}.
\end{equation}
\noindent The conditions (\ref{ref321-7e})-(\ref{ref321-7g}) are
defined taking into account the following properties of the
undisturbed velocity profile (\ref{ref31-12}): $[u_0]=0$,
$[u_0']=0$, $[u_0'']={\rm Gc}$, and $u_0(x_2)={\rm S}$.

The boundary value problem (\ref{ref321-7a})-(\ref{ref321-7g}) is
a spectral amplitude problem. The conditions of nontrivial
solution define perturbation growth rate as a function of the
parameters $\rm Gr$, $\rm Pr$, $\rm Gc$, $\rm S$ and $k$. The
solution with $\lambda_r=0$ determines the neutral behavior and
separates the regions of unstable modes with $\lambda_r>0$ from
those of stable modes with $\lambda_r<0$.

In the particular case of ${\rm Gc}=0$, the particles have no
influence on the fluid motion, and the boundary value problem
(\ref{ref321-7a})-(\ref{ref321-7g}) reduces to the stability
analysis \cite{gershuni-zhuhovitsky-76} of a flow with the odd
velocity profile (\ref{ref31-8}). For ${\rm Pr}=1$, the
instability appears above the critical value ${\rm
Gr}_{min}=496.3$, which is reached at $k_{min}=1.404$, and
corresponds to the monotonic (i.e. to the solution with zero
imaginary part of the growth rate $\lambda_i=0$) ``hydrodynamic''
perturbations.

In the case of ${\rm Gc}\ne 0$, the problem admits analytical
solution in the limit of short wavelength behavior for a small,
but finite width of the dust cloud. In the general case, the
boundary value problem (\ref{ref321-7a})-(\ref{ref321-7g}) was
treated numerically by the standard shooting and differential
sweep methods, which gave very close agreement of the results.

\subsubsection{Thin dust cloud, short wavelength limit}

In the case of a thin cloud, the parameter $d$ is small, and the
problem can be simplified. The nonlinear algebraic equations
(\ref{ref31-13a}), (\ref{ref31-13b}), defining the width of the
cloud $d$ and the coordinate of its center $x_c$, can be solved
explicitly with the accuracy $O\left(d^2\right)$:
\begin{equation}\label{ref322-1}
x_c=\frac{1}{\sqrt 3}, \quad d=\frac{1}{3 \sqrt{3}{\rm Gc}} \left(
{\rm Gr}-9\sqrt{3}{\rm S} \right).
\end{equation}
\noindent In the limit of vanishing cloud width, when $d \to 0$,
we obtain the lower threshold for the existence of a dust cloud
${\rm Gr}=9\sqrt 3{\rm S}$, which is in agreement with the the
value obtained above for an arbitrary value of $d$. It is clearly
seen from (\ref{ref31-12}) that in this case the velocity profile
coincides with (\ref{ref31-8}). A thin cloud is located in the
vicinity of the point at which the fluid velocity is maximal.

The short wavelength limit means that the wave number is large $k
\to \infty$. Let us investigate the case of small, but finite
values of $d$. Formally, this case corresponds to the limiting
transition, when $d \to 0$, $k \to \infty$, but their product $k
d$ is finite.

This particular case can be examined in the context of the
auxiliary problem, in which the layers of pure fluid are
semi-infinite and the dust cloud is in between these layers and
has the finite width $2d$. It is convenient to treat the problem
using the multi-scales technique.\cite{nayfeh-81} Let us introduce
the ``fast'' coordinate $\xi=(x-x_c)/\varepsilon$, where
$\varepsilon=1/k$ is the small parameter. In terms of the fast
coordinate the dust cloud occupies the finite region $-\Delta <
\xi < \Delta$ with the center at the point $\xi=0$. Here $\Delta =
k d$ is a semi-width of the cloud, measured in the units of the
fast coordinate. The layers of pure fluid occupy the areas
$-\infty < \xi < -\Delta$ and $\Delta < \xi < \infty$,
respectively.

The boundary value problem (\ref{ref321-7a})-(\ref{ref321-7g}) is
rewritten in terms of the fast coordinate $\xi$. As a result we
obtain the equations:
\begin{eqnarray}
\varepsilon^2\lambda \hat{u} -i\varepsilon u_0 \hat{u} &=&
-\varepsilon \hat{q}'+\hat{u}''-\hat{u}, \label{ref322-2a} \\
\varepsilon^2\lambda \hat{w} -i\varepsilon u_0 \hat{w} +
\varepsilon^2 u_0' \hat{u} &=& i\varepsilon
\hat{q}+\hat{w}''-\hat{w} \nonumber \\
&& +\varepsilon^2{\rm Gr}\hat{\vartheta},
\label{ref322-2b} \\
\hat{u}'-i\hat{w} & = & 0, \label{ref322-2c} \\
\varepsilon^2\lambda \hat{\vartheta} -i\varepsilon u_0
\hat{\vartheta} + \varepsilon^2 \hat{u} & = & \frac{1}{{\rm
Pr}}\left(\hat{\vartheta}''-\hat{\vartheta}\right),
\label{ref322-2d}
\end{eqnarray}
\noindent which are the same for all three zones: 1) $-\infty <
\xi < -\Delta$, 2) $-\Delta < \xi < \Delta$, 3) $\Delta < \xi <
\infty$. We require that the functions $\hat{u}$, $\hat{w}$,
$\hat{q}$, $\hat{\vartheta}$ must be bounded at $\xi=\pm \infty$.
The conditions at the borders of the cloud are as follows:
\begin{eqnarray}
\xi = \pm \Delta: && [\hat{q}]=0, \;\; [\hat{u}]=0, \;\;
[\hat{w}]=0, \nonumber \\
&& [\hat{w}']=-\varepsilon {\rm Gc}\hat{\zeta}_{1,2}, \;\;
[\hat{\vartheta}]=0, \;\;
[\hat{\vartheta}']=0, \label{ref322-2e} \\
\xi = -\Delta: && \left\{\varepsilon\lambda -i(u_0-{\rm
S})\right\}\hat{\zeta}_1=\varepsilon \hat{u}, \\
\xi = \Delta: && \lambda\hat{\zeta}_2=\hat{u}, \label{ref322-2f}
\end{eqnarray}
\noindent where $\hat{\zeta}_1$ and $\hat{\zeta}_2$ refer to the
points $\xi=-\Delta$ and $\xi=\Delta$, respectively.

For each zone the solution is sought as a power series in
$\varepsilon$. Taking into account (\ref{ref322-1}), we rewrite
the velocity profile (\ref{ref31-12}) with the accuracy
$O\left(\varepsilon^3\right)$:
\begin{eqnarray*}
u_0^{(1)} & = & {\rm S}-\varepsilon^2 \left\{ \frac{\rm Gr}{\sqrt
3}\frac{(\xi+\Delta)^2}{2}+\Delta\left(\frac{\rm Gr}{\sqrt 3}
-{\rm Gc}\right)\left(\xi+\Delta\right) \right\}, \\
u_0^{(2)} & = & {\rm S}-\varepsilon^2 \left(\frac{\rm Gr}{\sqrt 3}
-{\rm Gc}\right)\left( \frac {\xi^2-\Delta^2}{2}\right), \\
u_0^{(3)} & = & {\rm S}-\varepsilon^2 \left\{ \frac{\rm Gr}{\sqrt
3}\frac{(\xi-\Delta)^2}{2}-\Delta\left(\frac{\rm Gr}{\sqrt 3}
-{\rm Gc}\right)\left(\xi-\Delta\right) \right\}.
\end{eqnarray*}

The boundary value problem (\ref{ref322-2a})-(\ref{ref322-2f}) is
the inner one with respect to the initial problem
(\ref{ref321-7a})-(\ref{ref321-7g}), which in its turn should be
referred to as an outer one. Generally, the solutions of inner
problem is matched to the outer one. However, this should not
necessary be done in the present case since the solution of the
inner problem (\ref{ref322-2a})-(\ref{ref322-2f}) proves to be
vanishing at $\xi=\pm \infty$.

The solution of the boundary value problem
(\ref{ref322-2a})-(\ref{ref322-2f}) is quite cumbersome and is not
adduced here. Actually, the expressions for complex growth rates
are of primary interest. In the zero order we obtain the trivial
solution $\lambda^{(0)}=0$ and in higher orders we find the
nonzero corrections to $\lambda^{(0)}$, corresponding (with
accuracy up to the small terms of higher order) to two different
types of deformation of the dust cloud borders
$\hat{\zeta}_1=\pm\hat{\zeta}_2$:
\begin{equation}\label{ref322-3}
\lambda^{(1)}=\pm ikd\,{\rm Gc}\,{\rm e}^{-2kd}, \;\;
\lambda_r^{(2)}=\mp kd\,{\rm Gc\,S}\left(1+2 k d\right){\rm e}^{-2
k d}.
\end{equation}
One solution corresponds to deformation of the cloud borders in
the ``phase,'' when $\hat{\zeta}_1=\hat{\zeta}_2$. This solution
is stable since the real part of the growth rate is negative. In
the other solution, allowing for deformation of the cloud borders
in the ``anti-phase,'' when $\hat{\zeta}_1=-\hat{\zeta}_2$, the
real part of the growth rate is positive, and hence this solution
is unstable. In both solutions the imaginary part of the growth
rate is nonzero suggesting that the behavior of the system is
oscillatory.

Note, that the obtained result is true for any nontrivial values
the parameters $\rm Gc$ and $\rm S$. Thus, we conclude that the
dust cloud of a small, but finite width is unstable with respect
to short wavelength perturbations.

\subsubsection{Dust cloud of an arbitrary width}

Let us proceed to the discussion of general numerical results,
obtained for a cloud of arbitrary width $d$, and consider the case
${\rm S}=1$.
%
%
\begin{figure}[!t]
\includegraphics[width=0.35\textwidth]{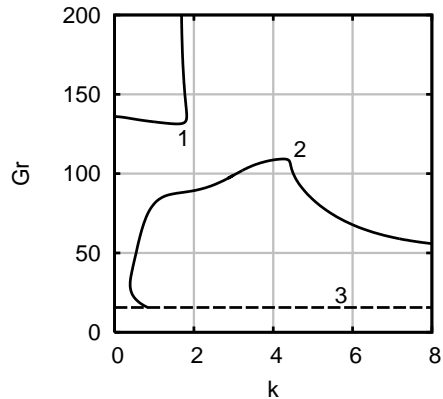}
\caption{Neutral curves (line 1 and 2), plotted for ${\rm
Gc}=100$, ${\rm S}=1$. Regions of unstable behavior are above the
line 1 and under the line 2. Line ${\rm Gr}=9\sqrt 3{\rm S}$ (line
3) corresponds to a cloud of the vanishing width.}
\label{fig:neutral_Gr_k}
\end{figure}
%
%

The neutral curves for a fixed value of the concentration
parameter ${\rm Gc}=100$ are presented in
Fig.~\ref{fig:neutral_Gr_k}. The instability regions are above the
curve $1$ (between the curve $1$ and the line $k=0$) and under the
curve $2$ (between the curves $2$ and $3$). Curve $3$ is a
straight line at which ${\rm Gr}=9\sqrt 3{\rm S}$. This line
defines the area of existence of a dust cloud and corresponds to a
dust cloud of zero width. The cloud exists in the domain that is
above this line. As can be seen, there is a narrow range of values
of $\rm Gr$, where the system is stable. The results
(\ref{ref322-3}), valid for a thin cloud in the short wavelength
limit, are in agreement with those obtained from numerical
solution to the general boundary value problem
(\ref{ref321-7a})-(\ref{ref321-7g}). However, it is obvious from
Fig.~\ref{fig:neutral_Gr_k}, that the short wavelength instability
is not ``the most dangerous'': although in some range of governing
parameters the flow is stable with respect to short waves, but yet
for all possible values of the cloud width $d$ there exist
unstable two-dimensional perturbations in the form of transversal
rolls with finite values of $k$. It is interesting to note, that
the steady state in the case of a thin cloud is unstable with
respect to perturbations with any wave number $k$ larger than some
critical value.

\begin{figure}[!t]
\includegraphics[width=0.35\textwidth]{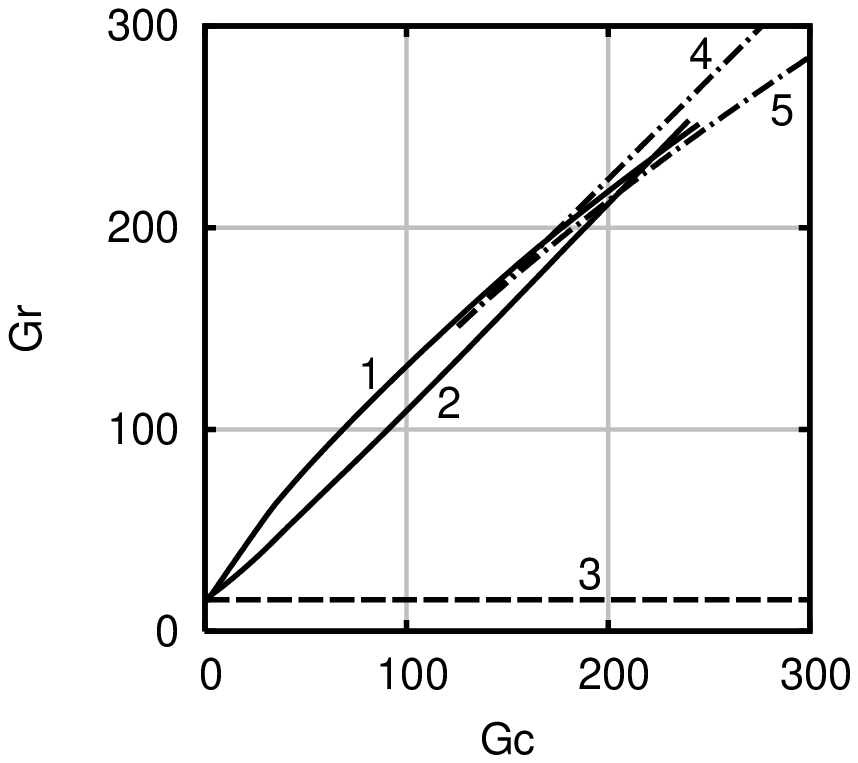}
\caption{Stability curves for the upper (line 1) and the lower
(line 2) branches of the dependence $d\left({\rm Gc}\right)$ on a
plane ($\rm Gc$, $\rm Gr$) at ${\rm S}=1$. The stability regions
are under the line 1 and above the line 2. Line 3 is the line
${\rm Gr}=9\sqrt 3{\rm S}$; lines 4 and 5 represent bounds of the
hysteresis zone (see Fig.~\ref{fig:wedge_Gr_Gc}).}
\label{fig:stab_S1}
\end{figure}

The global minimum of the curve $1$ and the global maximum of the
curve $2$ over $k\ge0$ in Fig.~\ref{fig:neutral_Gr_k} are
characterized by the critical values of the Grashof number: ${\rm
Gr}_{min}$, ${\rm Gr}_{max}$, which are reached at the values of
the wave number $k_{min}$, $k_{max}$ respectively. Let us discuss
the dependence of these critical parameters on the concentration
parameter $\rm Gc$. Figure~\ref{fig:stab_S1} gives the stability
diagram on the plane ($\rm Gc$, $\rm Gr$). The stability region is
under the line $1$ and above the line $2$. Note, that the curves
$1$ and $2$ refer to the upper and the lower branches of the
dependence $d\left({\rm Gc}\right)$, respectively (see
Fig.~\ref{fig:hyster_d_Gc}). This becomes important for
distinguishing between stability regions for the upper and lower
branches in the hysteresis zone (Fig.~\ref{fig:wedge_Gr_Gc}),
bounded by lines 4 and 5: the stability region of the upper branch
is defined by the curve 1, for the lower branch it is defined by
the curve 2. As before, the line $3$ is a line determining the
region of a cloud of vanishing width, at which ${\rm Gr}=9\sqrt
3{\rm S}$.
%
%
%
%

It should be noticed that the section of the straight line ${\rm
Gc}=0$, at $9\sqrt 3 < {\rm Gr} < 496.3$ also refers to a
stability region. A piece of this line, corresponding to the
values $0 \le {\rm Gr} < 9\sqrt 3$ is stable as well, however, the
basic steady state in this case is completely free of particles,
i.e. a dust cloud is absent.

\begin{figure}[!b]
\includegraphics[width=0.38\textwidth]{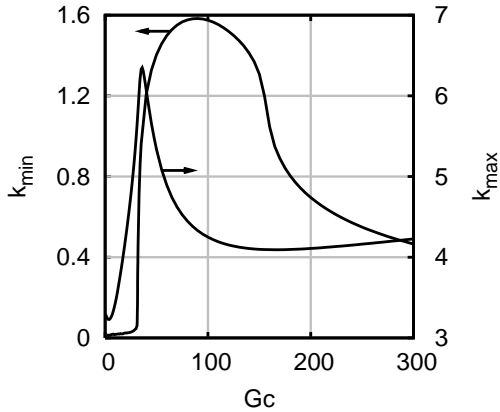}
\caption{Critical wave numbers $k_{min}$, $k_{max}$ as functions
of ${\rm Gc}$ at ${\rm S}=1$.} \label{fig:k_crit}
\end{figure}

The variation of  $k_{min}$, $k_{max}$ with the concentration
parameter $\rm Gc$ is shown in Fig.~\ref{fig:k_crit}. The
imaginary parts of growth rates corresponding to these solutions
are nonzero, which means that this ``concentration'' mode is
oscillatory. The hydrodynamic perturbations are not the most
dangerous for our particular value of sedimentation parameter $\rm
S$, their stability bound lies much higher than the lines $1$ and
$2$. Of special note is the dependence $k_{min}({\rm Gc})$
 at relatively small values of the concentration parameter
$\rm Gc$. It is seen, that for not very large values of $\rm Gc$
the wave number $k_{min}$ is rather small, i.e. the system
demonstrates a nearly long wavelength behavior.
%
%
%
%

\begin{figure}[!t]
\includegraphics[width=0.35\textwidth]{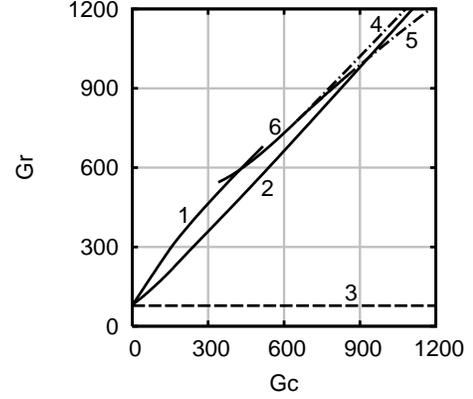}
\caption{Stability curves for the upper (lines 1 and 6) and the
lower (line 2) branches of $d\left({\rm Gc}\right)$ on a plane
($\rm Gc$, $\rm Gr$). The stability regions are under the line 1
and 6 and above the line 2. Line 3 is the line ${\rm Gr}=9\sqrt
3{\rm S}$; lines 4 and 5 represent bounds of the hysteresis zone.
} \label{fig:stab_S5}
\end{figure}

With the increase of the sedimentation parameter $\rm S$ the
region of hysteresis wedge shifts to the region of larger values
of $\rm Gr$ and $\rm Gc$. Fig.~\ref{fig:stab_S5} presents the
stability diagram for the case ${\rm S}=5$. The stability curves
$1$ and $2$ do not undergo any qualitative changes. However, in
contrast to the case ${\rm S}=1$, the concentration mode is in
competition with the hydrodynamic mode: there appears the range of
values of $\rm Gc$, at which the hydrodynamic perturbations (line
6) become the most dangerous for the upper branch of the
dependence $d({\rm Gc})$. With increase of $\rm S$ this range
grows, reducing the total stability region. In the hysteresis zone
line 6 sticks to the line 5: the upper branch becomes unstable
here. In comparison with the case ${\rm Gc}=0$, the hydrodynamic
mode is no longer monotonic. Indeed, the monotonic character of
hydrodynamic perturbations is specified
\cite{gershuni-zhuhovitsky-76} by the symmetry of the velocity
profile $u_0$. In our case, as it follows from (\ref{ref31-12}),
the symmetry of $u_0$ is broken at any value ${\rm Gc}\ne0$.
%
%
%
%

\subsubsection{Arbitrary three-dimensional perturbations}

It was mentioned before, that after linearization
Eqs.~(\ref{ref321-4a})-(\ref{ref321-4c}), boundary conditions
(\ref{ref321-5}) and the conditions at the borders of the dust
cloud (\ref{ref321-2}) admit a transformation, which is analogous
to the Squire transformation.\cite{squire-33} Under such a
transformation the Prandtl number $\rm Pr$ does not change,
whereas the parameters $\rm Gr$, $\rm Gc$, and $\rm S$ transform
according to the following rule:
\begin{equation} \label{ref324-1}
{\rm Gr}=\frac{k_z}{k}{\rm Gr'}, \quad {\rm Gc}=\frac{k_z}{k}{\rm
Gc'}, \quad {\rm S}=\frac{k_z}{k}{\rm S'},
\end{equation}
\noindent where $\rm Gr$, $\rm Gc$, $\rm S$ are the parameters of
the two-dimensional problem and $\rm Gr'$, $\rm Gc'$, $\rm S'$ are
the corresponding parameters of the full three-dimensional
problem; $k_y$ and $k_z$ are the wave numbers along the axes $y$
and $z$, respectively; $k^2=k_y^2+k_z^2$. Note, that at any given
value $k_z/k$ the parameters of the two-dimensional problem $\rm
Gr$, $\rm Gc$, $\rm S$ are always less than those of the
three-dimensional problem. However, this fact does not imply that
two-dimensional perturbations are the most dangerous.

Let us introduce the parameter, describing deviation of arbitrary
three-dimensional perturbations from the two-dimensional
perturbations in the form of transverse rolls (\ref{ref321-6}),
namely the angle between the wave vector ${\bf k}=(0,k_y,k_z)$,
which generally lies in the plane $y$-$z$, and the $z$-axis. If we
denote the cosine of this angle by $\alpha$, then it follows from
(\ref{ref324-1})
$$
{\rm Gr}'=\frac{\rm Gr}{\alpha}, \quad {\rm Gc}'=\frac{\rm
Gc}{\alpha}, \quad {\rm S}'=\frac{\rm S}{\alpha}.
$$
The greatest distinction of arbitrary three-dimensional
perturbations from transversal rolls (\ref{ref321-6}) corresponds
to the limiting case of longitudinal rolls, when $\alpha=0$
$(k_z=0$, $k\ne 0)$. The perturbations of this kind are often
called the ``helical'' perturbations, since the trajectory of a
fluid element in such a flow looks like a helix. The motion of a
fluid particle can be considered as a sum of two different forms
of motion: first, the particle moves circle-wise inside the
longitudinal roll, second, it is carried along the roll with the
velocity $u_0+u$.

Thus, at given $\alpha$ the diagram of stability with respect to
arbitrary three-dimensional perturbations can be obtained from
that of two-dimensional perturbations (\ref{ref321-6}) by
rescaling $\rm Gr$ and $\rm Gc$ by a factor of $1/\alpha$. It is
important to note, that the parameter $\rm S$ is also transformed.

The linear stability analysis, performed with respect to
two-dimensional perturbations (\ref{ref321-6}) for $\rm S$ from
$1$ up to $5$, allows us to investigate the influence of
three-dimensional perturbations with $\alpha$ in the range from
$0.2$ up to $1$ for ${\rm S}=5$. It turns out, that in this range
the most dangerous are the perturbations in the form of
transversal rolls.

Further, it can be explicitly shown, that the limiting case of the
longitudinal rolls (helical perturbations), when $\alpha=0$
$(k_z=0$, $k\ne 0)$, does not cause instability. Indeed, as it
follows from Eqs.~(\ref{ref321-4a})-(\ref{ref321-5}) and
linearized form of conditions (\ref{ref321-2}), (\ref{ref321-3}),
this particular case is described by the following boundary value
problem (the equations are the same for each of three zones):
\begin{eqnarray}
\lambda \hat{u} & = & -\hat{q}'+\hat{u}''-k_y^2\hat{u}, \label{ref324-2a} \\
\lambda \hat{v} & = & ik_y\hat{q}+\hat{v}''-k_y^2\hat{v}, \label{ref324-2b} \\
\hat{u}'-ik_y\hat{v} & = & 0, \label{ref324-2c} \\
\lambda \hat{w} + u_0'\hat{u} & = & \hat{w}''-k_y^2\hat{w}+{\rm
Gr}\hat{\vartheta}, \label{ref324-2d}
\\
\lambda\hat{\vartheta} +\hat{u} & = & \frac{1}{\rm
Pr}\left(\hat{\vartheta}''-k_y^2\hat{\vartheta}\right),
\label{ref324-2e}
\end{eqnarray}
\begin{eqnarray}
x = \pm 1: && \hat{u}=0, \;\; \hat{v}=0, \;\; \hat{w}=0, \;\;
\hat{\vartheta}=0, \label{ref324-2f} \\
x = x_{1,2}: && [\hat{q}]=0, \;\; [\hat{u}]=0, \;\; [\hat{v}]=0,
\;\;  [\hat{\vartheta}]=0, \nonumber \\
&&[\hat{\vartheta}']=0, \;\; [\hat{w}]=0, \;\; [\hat{w}']=-{\rm
Gc}\,\hat{\zeta}_{1,2}, \nonumber \\
&&\lambda\hat{\zeta}_{1,2}=\hat{u}. \label{ref324-2g}
\end{eqnarray}

The problem for the fields $\hat{u}$ and $\hat{v}$ splits off and
can be treated independently. Excluding the pressure $\hat{q}$
from Eqs.~(\ref{ref324-2a})-(\ref{ref324-2c}) and taking into
account the continuity of $\hat{u}$ and $\hat{v}$ at the points
$x_1$ and $x_2$, we obtain the following boundary value problem:
\begin{eqnarray}
\lambda(\hat{v}''-k_y^2\hat{v}) & = &
\hat{v}^{IV}-2k_y^2\hat{v}''+k_y^4\hat{v},
\label{ref324-3a} \\
x = \pm 1: \;\; \hat{v} & = & 0, \;\; \hat{v}'=0.
\label{ref324-3b}
\end{eqnarray}
\noindent Let us multiply Eq.~(\ref{ref324-3a}) by the complex
conjugate $\hat{v}^{*}$ and integrate by parts across the layer.
Taking into account (\ref{ref324-3b}), after elementary
calculations we obtain
$$
\lambda=-\frac{\left< \left| \hat{v}'' \right|^2 +2k_y^2 \left|
\hat{v}' \right|^2+k_y^4\left| \hat{v} \right|^2\right>} {\left<
\left| \hat{v}' \right|^2 +k_y^2 \left| \hat{v}
\right|^2\right>}<0,
$$
\noindent where $\left<\ldots\right>=\int_{-1}^{1} \ldots \,dx$.
Since the problem (\ref{ref324-3a})-(\ref{ref324-3b}) describes
perturbations in a quiescent viscous uniform fluid, it cannot give
birth to instability: the growth rate is proved to be real and
negatively defined; the perturbations monotonically decrease with
time, and, therefore, one should look for a mode with $\hat{v}=0$.
Hence, with account of incompressibility relation
(\ref{ref324-2c}), the boundary conditions (\ref{ref324-2f}) for
$\hat{u}$ and continuity of this field at the points $x_1$ and
$x_2$ we obtain $\hat{u}=0$.

Since $\hat{u}=0$, equation (\ref{ref324-2e}) corresponds to
diffusion of $\hat{\vartheta}$ in motionless fluid, and therefore,
from analogous consideration we conclude that perturbations
decrease. Hence, the mode with $\hat{\vartheta}=0$ is to be found.
As it follows from the condition (\ref{ref324-2g}) for
$\hat{\zeta_1}$ and $\hat{\zeta_2}$, the solution with $\lambda
\ne 0$ is possible only if $\hat{\zeta_1}=0$ and
$\hat{\zeta_2}=0$. Consequently, $\hat{w}$ and its derivative are
continuous at the points $x_1$, $x_2$, and the problem for
$\hat{w}$ coincides with the problem for $\hat{\vartheta}$ with
the rescaled time $t \,{\rm Pr}^{-1}$. Thus, for any values of
governing parameters the basic state is proved to be stable with
respect to perturbations in the form of longitudinal rolls.

\section{Conclusions}

The interaction of the vortex buoyancy convective flow laden with
particles of dust has been investigated in the case when the
volume concentration of the solid phase is small. Under these
conditions the particles are partially carried away by the flow,
but due to sedimentation their velocity differs from the velocity
of a fluid. At sufficiently intensive thermal convection some
portion of the particles can be captured by the flow, which
eventually results in formation of a cloud of dust.

If mass concentration of the particles is rather small, the
particles do not influence the flow and the study of the formation
of a dust cloud is reduced to a kinematic problem. The size of the
dust cloud captured by the convective vortex monotonically
increases with the growth of the thermal Grashof number from some
threshold value ${\rm Gr_*}$. At the values ${{\rm Gr} < {\rm
Gr}_*}$, the formation of a steady cloud is impossible: sooner or
later all the particles settle on the bottom of a cavity. This
critical value of the Grashof number is determined by the cavity
shape and thermal boundary conditions and is also proportional to
the sedimentation parameter ${\rm S}$. The calculations for the
case of an infinite vertical layer heated from the sidewalls give
${\rm Gr}_* = 9 \sqrt{3} {\rm S}$.

With the increase of mass concentration of the particles their
influence on the flow becomes significant. This happens when
relative variations of mass concentration are comparable with the
Boussinesq nonisothermality parameter $\beta\theta$. Generally,
the growth of mass concentration results in the decrease of flow
intensity. Energy of the flow is partly consumed on the particle
motion against the gravity and on the motion of particles together
with the descending flow. This energy is not returned to the flow
completely because of the dissipative loss. The calculations show
that the interaction of the particles with the convective flow can
lead to the hysteresis phenomena. Different flow patterns can
arise at the same values of the governing parameters: the modes,
in which a comparatively small dust cloud strongly suppresses the
ascending flow and the modes developing at comparatively large
cloud when the density inhomogeneities are low and the suppression
effect is insignificant.

The linear stability of the basic steady state in the form of a
dust cloud in an infinite layer, heated from sidewalls, has been
investigated. It is shown, that in a relatively narrow range of
the values of governing parameters this state is stable. Based on
the results of the stability analysis, we can infer that in the
typical case of a closed cavity the steady state with a relatively
large dust cloud is most likely to have a larger range of
stability. In the case of an infinite layer such a steady state is
broken by long wavelength perturbations. Obviously, such
perturbations cannot arise in a closed container of a finite size.

The performed investigation demonstrates that the described
effects of two-way interaction of fluid and particles can exist in
natural environment or be realized experimentally. The values of
the sedimentation parameter ${\rm S}$ of order $1$, used in the
linear stability analysis, correspond, for example, to the case of
rather fine particles of dust with $r_p\sim 10^{-4}~{\rm cm}$,
suspended under gravity in a gaseous medium filling a
laboratory-scale container $L \sim 1~{\rm cm}$ ($r_p/L \sim
10^{-4}$, $\delta \sim 10^3$, ${\rm Ga} \sim 10^5$). The developed
theory is quite general and can be applied to describe similar
effects not only in dusty media, but also in aerosols, liquids
laden with small solid particles, and biological species in
aqueous media.

\section{Acknowledgments}

The research was partially supported by INTAS (Grant No.
2000-0617) and CRDF (Grant No. PE-009-0), which are gratefully
acknowledged.


\end{document}